# Superconducting versus normal conducting cavities

*Holger Podlech*
Institute for Applied Physics (IAP), Frankfurt am Main, Germany

**Abstract**
One of the most important issues of high-power hadron linacs is the choice of technology with respect to superconducting or room-temperature operation. The favour for a specific technology depends on several parameters such as the beam energy, beam current, beam power and duty factor. This contribution gives an overview of the comparison between superconducting and normal conducting cavities. This includes basic radio-frequency (RF) parameters, design criteria, limitations, required RF and plug power as well as case studies.

## 1  Introduction

Worldwide there is an increasing interest in a new generation of high-power proton and ion linacs. The term 'high power' refers to the product of beam energy and beam current which is the beam power. Typical applications are neutron production with low-energy deuterons, neutron production using high-energy protons, nuclear physics at rare isotope beam facilities or nuclear waste transmutation. The typical modern hadron linac consists of three major parts (Fig. 1):

1. front end (low-energy part);
2. drift tube linac (intermediate-energy part);
3. elliptical section (high-energy part).

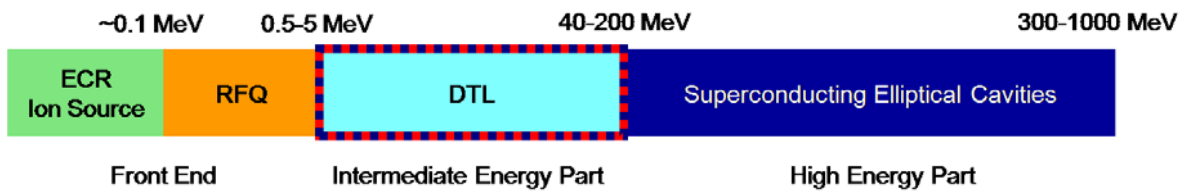

**Fig. 1:** Scheme of a modern high-power hadron linac

In most cases the front-end consists of an electron cyclotron resonance (ECR) ion source, the low-energy beam transport (LEBT) and a radio-frequency quadrupole (RFQ) as first accelerating RF structure. The vast majority of all RFQ structures have been realized using normal conducting technology. The high-energy section starts typically between 100 AMeV and 200 AMeV and mostly makes use of superconducting elliptical multicell cavities. Even for lower duty factors superconducting elliptical cavities are a good choice with respect to the overall required radio-frequency (RF) and plug power. In the case of the intermediate-energy part, the situation is not as clear. The choice of technology depends on the beam current, acceleration gradients and most importantly on the duty factor. Figure 1 shows schematically a modern high-power hadron linac.

The Spallation Neutron Source SNS (ORNL, USA) [1] which has been taken into operation in recent years is considered as the prototype for a new generation of high-power linacs. The driver linac provides 1 GeV protons. The duty factor is 6 % with a peak current of 38 mA resulting in an average beam current of 1.4 mA and a beam power of 1.4 MW. The front end accelerates the beam to 2.5 MeV. The intermediate-energy section consists of a classical Alvarez DTL (2.5 MeV to 87 MeV,

402.5 MHz), followed by a normal conducting CCL linac (87 MeV to 180 MeV, 805 MHz). The high-energy section consists of two groups of superconducting elliptical five-cell cavities operated at 805 MHz. The transition energy between normal conducting and superconducting cavities is 180 MeV. There are other projects such as MYRRHA [2] which will deliver a continuous wave (cw) proton beam. For this high duty factor, a much lower transition energy of 3.5 MeV has been chosen. Table 1 summarizes basic parameters of different hadron linacs which are in operation, under construction or in the design phase [1–10].

**Table 1:** Parameters of different modern hadron linacs

| Project | Particles | Final energy (AMeV) | Transition energy (AMeV) | Duty factor (%) | Pulse beam current (mA) |
|---|---|---|---|---|---|
| SNS | p | 1000 | 187 | 6 | 38 |
| ESS | p | 2500 | 50 | 4 | 50 |
| MYRRHA | p | 600 | 3.5 | 100 | 4 |
| IFMIF | d | 20 | 2.5 | 100 | 125 |
| SPIRAL2 | d, HI | 20 | 0.75 | 100 | 5 |
| FRIB | p-HI | 200–600 | 0.3 | 100 | <1 |
| SARAF | p, d | 20 | 1.5 | 100 | 2 |
| LINAC4/SPL | p | 5000 | 160 | <10 | 40 |
| FAIR p-Linac | p | 70 | 70 | 0.04 | 70 |
| GSI SHE | HI | 7.5 | 1.4 | 100 | 1 |

p, protons; d, deuterons; HI, heavy ions.

Figure 2 (left) shows the transition energy as a function of the duty factor. In general, it can be observed that the transition energy is lower at higher duty factors because superconducting operation more likely becomes advantageous. There is also a clear dependence between the transition energy and the beam current (Fig. 2, right). The higher the pulse beam current the higher the transition energy typically is.

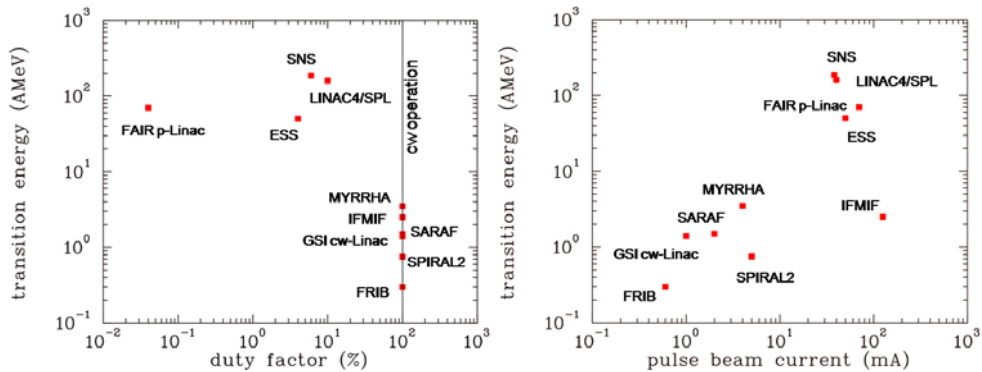

**Fig. 2:** Transition energy between normal conducting and superconducting technology for different duty factors (left) and for different pulse beam currents (right).

## 2   RF parameters

This section gives a brief description of different RF parameters. In general, these parameters can be divided into two groups:

  1. parameters dependent on the surface resistance;

  2. parameters independent of the surface resistance.

The surface-independent RF parameters are mostly used to compare different cavity geometries independent of the preferred technology.

### 2.1   Surface resistance $R_s$

The presence of RF fields results in a surface resistance $R_s$ in both cases. The physical reason for this resistance and its magnitude is different, however, for normal conducting and superconducting cavities.

In the case of normal conducting cavities the skin effect leads to an expulsion of the electromagnetic fields and the corresponding current density. The decrease of the current density can be described by the following expression:

$$j(x) = j_0 e^{-x/\delta} \cos(x/\delta) \tag{1}$$

$\delta$ is the so-called "skin depth" which is the equivalent thickness of a homogeneous current density $j_0$. The skin depth is inversely proportional to the conductivity $\sigma$ and inversely proportional to the square root of the frequency. Finally, the surface resistance can be calculated:

$$R_s = \frac{1}{\sigma \delta} = \sqrt{\frac{\pi \mu_0 \mu_r f}{\sigma}} \tag{2}$$

Typical values for the surface resistance of normal conducting cavities are several milliohms.

In the case of superconducting cavities, the Meissner–Ochsenfeld effect leads to an almost complete expulsion of the electromagnetic fields from the superconductor. The current decays exponentially from the surface and reaches $1/e$ at the London penetration depth which is 39 nm for niobium. In the case of static fields, the current transport in superconductors is loss free but not for RF fields. Owing to the inertia of Cooper pairs unpaired electrons are then not completely shielded from the time-varying fields. These electrons close to the Fermi edge can be accelerated by the RF fields within the penetration depth leading to a non-vanishing surface resistance. The surface resistance can be expressed by the Bardeen–Cooper–Schrieffer (BCS) theory, for example for niobium:

$$R_s(BCS) = 2 \cdot 10^4 \frac{1}{T} \left(\frac{f}{1.5}\right)^2 e^{-\frac{17.67}{T}} \tag{3}$$

The surface resistance is highly dependent on the temperature. It decreases exponentially with lower temperatures because the number of unpaired electrons is also decreasing. On the other hand, the resistance increases quadratically with the frequency. As a consequence 2 K is chosen as the operation temperature at higher frequencies instead of 4 K. For the standard material niobium typical values for the BCS surface resistance are of the order of several nanoohms to several tens of nanoohms. The surface resistance of superconducting cavities is typically five orders of magnitude lower than for normal conducting cavities. The surface resistance of superconducting cavities is, however, higher in reality because of a frequency-independent residual resistance and because of eventually trapped magnetic flux.

## 2.2 Dissipated power $P_c$

The surface resistance results in a dissipation of energy and leads consequently to a power density

$$\rho = \frac{1}{2} R_s |H|^2 \tag{4}$$

The total power $P_c$ can be obtained by integrating over the whole cavity surface:

$$P_c = \frac{1}{2} R_s \int_A |H|^2 \, dA \tag{5}$$

The power losses are several orders of magnitude lower for superconducting cavities because of the surface resistance dependency.

## 2.3 $Q$ value

The (unloaded or intrinsic) $Q$ value or quality factor of a cavity can be calculated by

$$Q_0 = \frac{\omega W}{P_c} \tag{6}$$

The stored energy $W$ can be calculated either using the electric or the magnetic field. On average the energy is equally distributed in both fields. Of course the stored energy is independent of the surface resistance. It depends only on the cavity geometry and the field level:

$$W = \frac{1}{2} \mu_0 \int_V |H|^2 \, dV = \frac{1}{2} \varepsilon_0 \int_V |E|^2 \, dV \tag{7}$$

Aside from a factor $2\pi$ the $Q$ value gives the number of RF periods until the stored energy is dissipated after the RF has been turned off. In addition, the $Q$ value measures the full width of the resonance curve where the field amplitude reaches $1/\sqrt{2}$ of the maximum value at resonance:

$$Q_0 = \frac{f}{\Delta f} \tag{8}$$

The unloaded $Q$ value depends inversely proportionally on the surface resistance. Typical $Q$ values for normal conducting cavities are between $10^3$ and $10^5$ and between $10^7$ and $10^{11}$ for superconducting cavities.

## 2.4 Shunt impedance $R_a$

A cavity can be described as an equivalent RCL parallel circuit. The shunt impedance $R_a$ is the real part of the frequency-dependent complex impedance $Z(\omega)$. At resonance, the $Z(\omega)$ and $R_a$ become identical. The shunt impedance describes the ability of the cavity to convert RF power into voltage:

$$R_a = \frac{U_a^2}{P_c} \tag{9}$$

Here $U_a$ is the effective or accelerating voltage including the transit time factor. Sometimes the shunt impedance is normalized to the length to compare cavities with different length:

$$Z_{eff} = \frac{R_a}{L} = \frac{U_a^2}{P_c L} \tag{10}$$

## 2.5 $R_s$-independent parameters

To compare different geometries it is helpful to use parameters which are independent of the surface resistance. The so-called geometrical factor $G$ is defined by

$$G = R_s Q_0 = \frac{R_s \omega W}{P_c} = \frac{\omega \mu_0 \int_V |H|^2 \, dV}{\int_A |H|^2 \, dA} \tag{11}$$

For a given frequency and field distribution it is the ratio between the cavity volume and cavity surface.

The shunt impedance is one of the most important parameters because it describes the cavity efficiency. This only works if the surface resistance is known. We can use the $R/Q$ value which is independent of the surface resistance to compare different RF structures with a given surface resistance:

$$\frac{R_a}{Q_0} = \frac{U_a^2}{\omega W} = \frac{2[\int E_z \cos(\omega z/\beta c) \, dz]^2}{\varepsilon_0 \omega \int_V |E|^2 \, dV} \tag{12}$$

The $R/Q$ value is also known as geometrical shunt impedance because it measures the ability of the cavity to focus the electric field on axis.

## 2.6 Example: the pillbox cavity

The pillbox cavity is a simple cylindrical cavity. The fundamental mode is the $TM_{010}$ mode which is also used in elliptical cavities. All RF parameters of the pillbox cavity can be calculated analytically. Table 2 shows the basic RF parameters and a comparison of a normal conducting and a superconducting pillbox cavity.

Table 2: Comparison of a normal conducting and a superconducting pillbox cavity

| Parameter | Normal conducting | Superconducting |
|---|---|---|
| Length (cm) | 10 | 10 |
| Radius (cm) | 7.65 | 7.65 |
| Frequency (MHz) | 1500 | 1500 |
| $U_a$ (MV) | 1 | 1 |
| $T$ (K) | 300 | 2 |
| $R_s$ ($\Omega$) | 0.01 | $2 \times 10^{-8}$ |
| $Q_0$ | 25 500 | $1.3 \times 10^{10}$ |
| $R_a$ ($\Omega$) | $5 \times 10^6$ | $2.5 \times 10^{12}$ |
| $W$ (J) | 0.54 | 0.54 |
| $P_c$ (W) | 198 000 | 0.4 |
| $G$ ($\Omega$) | 257 | 257 |
| $R/Q$ ($\Omega$) | 196 | 196 |

# 3 General consideration of power consumption

In this section the RF power and the grid power requirements of normal conducting and superconducting cavities are investigated. For a given RF structure the RF power is dominated by the following parameters:

1. accelerating gradient $E_a$;
2. beam current;
3. shunt impedance.

The required grid power is typically significantly higher than the RF power. It can be a crucial factor for operational costs especially for high duty factor accelerators. The required plug power to operate RF cavities is dominated by:

1. accelerating gradient $E_a$;
2. beam current;
3. shunt impedance;
4. duty factor;
5. RF amplifier efficiency;
6. efficiency of the cryogenic system (in a superconducting system);

The following considerations are limited to the cavity operation. The additional required power for magnets, heating, ventilation, etc., is not included. The given values for cavity parameters are only examples. In reality they can be higher or lower depending on the specific case and kind of RF structure.

Figure 3 shows a comparison of the required RF power (cavity loss $P_c$) without beam for a normal conducting and a superconducting cavity. The assumed surface resistance of the superconducting cavity is the BCS value at $T = 4$ K (12.6 n$\Omega$).

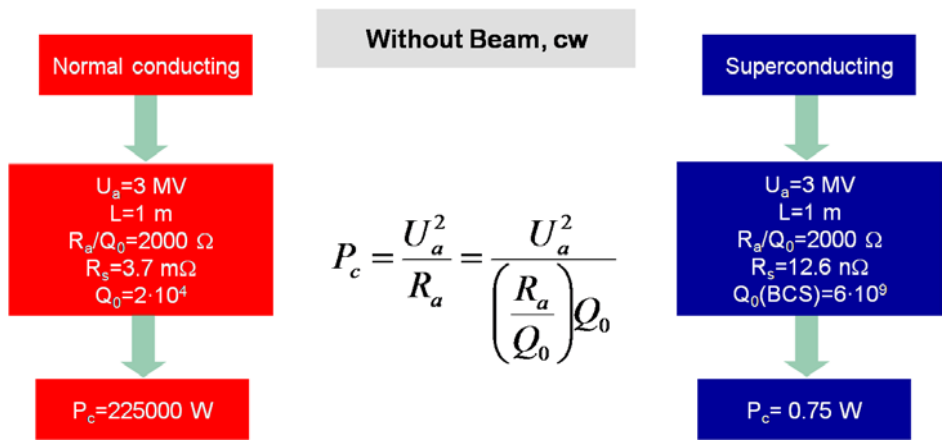

**Fig. 3:** Required RF power without beam for a normal conducting (left) and for a superconducting cavity (right)

In this example the difference in RF power is only due to the difference in the surface resistance. For superconducting cavities this is too optimistic because the real resistance can be significant higher. In addition to the BCS-value possible trapped magnetic flux and a frequency independent residual resistance will lead to higher power losses. The normal conducting cavity would require 225 kW/m. In case of 100 % duty factor this corresponds to the thermal load. The present technological limit is about 100 kW/m. To reach this value the gradient has to be reduced in reality from 3 MV/m to 2 MV/m. Figure 4 shows a comparison of the grid power. For the normal conducting cavity we have to take the efficiency of the RF amplifier into account which is typically 60 %. In addition to the RF losses we have an additional heat entry into the helium bath due to static losses in the case of superconductivity. The sum of dynamic and static losses have to be removed using a cryogenic system. Owing to the very low thermodynamic efficiency η of a cryogenic system operated at 4 K given by

$$\eta = \left(\frac{4K}{300K - 4K}\right) \cdot 0.25 = 0.003$$

the grid power is much larger than the heat load in the helium bath. Finally the ratio between the grid power for a normal conducting and a superconducting cavity has been decreased from more than 4 orders of magnitude to less than a factor 100 (cw, no beam).

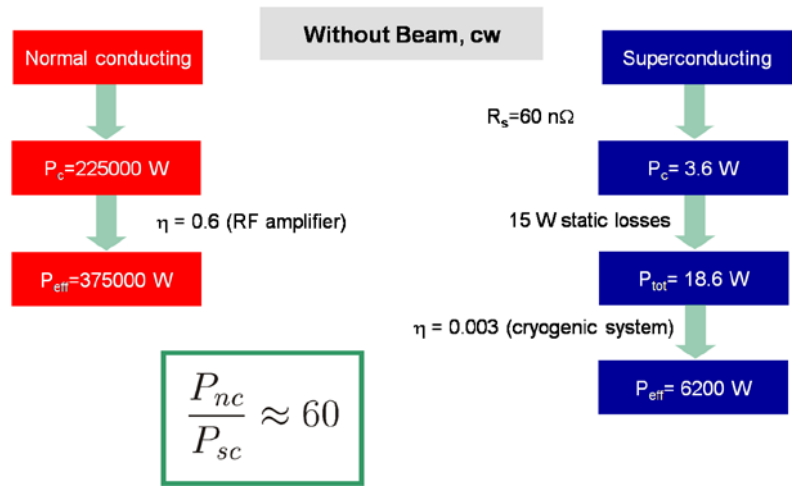

**Fig. 4:** Required grid power (cw operated) without beam for a normal conducting (left) and for a superconducting cavity (right)

The situation changes again in the presence of beam loading. Assuming a beam current of 20 mA we have 60 kW beam power per cavity ($U_a$ = 3 MV). In this example the normal conducting cavity would require 475 kW and the superconducting cavity 106 kW grid power (see Fig. 5). The required grid power of superconducting cavities with heavy beam load and high duty factor is dominated by the beam power.

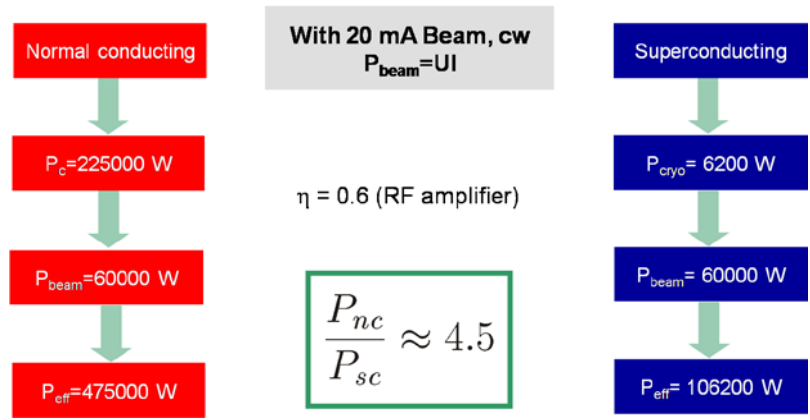

**Fig. 5:** Required grid power with beam (cw operated) for a normal conducting (left) and for a superconducting cavity (right)

In the case of higher beam currents and low duty factor the situation becomes more favourable for normal conducting cavities. Figure 6 shows an example with 20 mA current and a duty factor of 1 %. Owing to the static losses which are always present, the grid power is lower for the operation of normal conducting cavities.

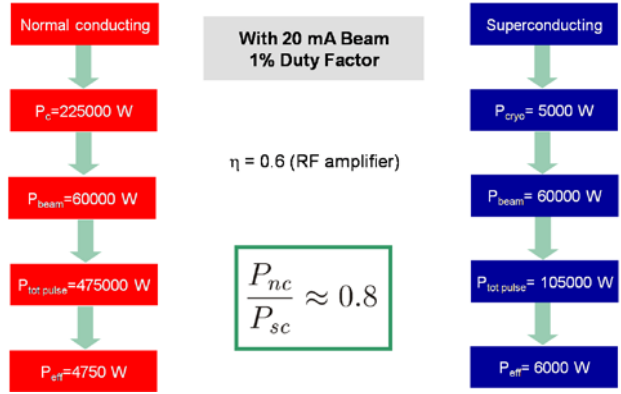

**Fig. 6:** Required grid power with beam and low duty factor (1 %) for a normal conducting (left) and for a superconducting cavity (right)

## 4    Systematic investigation of the required power

The topic of this section is a more systematic description of the required RF and grid power. The power has been calculated dependent on the gradient, duty factor, beam current, surface resistance and static losses. In each plot always one normal conducting and one superconducting cavity with typical parameters have been compared. The lengths of the cavities have been fixed to 1 m. A frequency of 325 MHz has been used. The assumed shunt impedance of the normal conducting cavity is 60 MΩ/m and the $R/Q$ value of the superconducting cavity is 3000 Ω. Of course, shunt impedance and surface resistance change with frequency, but the general results are very similar for different cavity shapes, frequencies and the β of the cavities.

In addition, the following assumptions have been made for reasons of simplicity.

1. No additional power for over-coupling of superconducting cavities.
2. No non-Ohmic effects (field emission) are considered.
3. The operation temperature of the helium bath is 4.2 K.
4. No power for auxiliary systems.
5. No cable losses.
6. Perfect match to the beam (no reflected power).
7. Duty factor of the beam and RF is equal.

Figure 7 shows the required RF power without beam as a function of the accelerating gradient for 1 m long normal conducting and superconducting RF structures with a β of 0.2. Only Ohmic losses are considered which results in a quadratic increase of the RF power with the accelerating gradient. The horizontal line represents a power level (thermal load) of 100 kW/m which is close to the present technological limit of cooling capabilities of normal conducting RF structures. This means that for this specific case the gradient is limited to about 2.5 MV/m for cw operation. Owing to the much lower surface resistance of superconducting cavities the required power is, according to this, lower by several orders of magnitude. The different curves take various values of the residual resistance $R_0$ into account which has to be added to the BCS value.

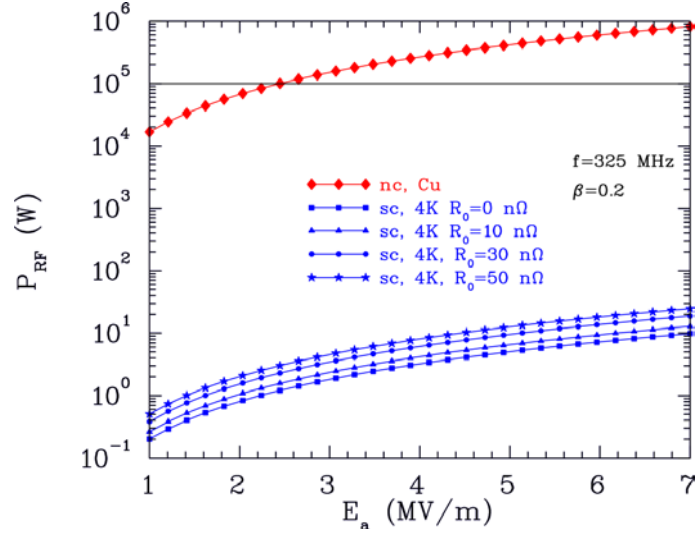

**Fig. 7:** RF power without beam as a function of the accelerating gradient $E_a$. Normal conducting cavity: $f$ = 325 MHz, $Z_{eff}$ = 60 MΩ/m, $\beta$ = 0.2, $L$ = 1 m. Superconducting cavity: $f$ = 325 MHz, $\beta$ = 0.2, $R/Q$ = 3000 Ω, $T$ = 4.2 K, $G$ = 55 Ω.

Figure 8 shows the required plug power for a given gradient without any beam load and 100 % duty cycle. In the case of normal conducting operation an efficiency of the RF amplifier of 60 % has been assumed. The static losses for the superconducting cavities are 5 W. Owing to the low thermo-dynamical efficiency of the cryogenic system, the ratio between normal conducting and superconducting operation is now only between one and two orders of magnitude. At very low gradients the plug power for superconducting operation is dominated by the static losses. Beam current requires additional RF power. Figure 9 shows an example with a gradient of 4 MV/m. For each 1 mA beam current, 4 kW additional RF power has to be provided. For normal conducting cavities the power losses are still dominant up to several tens of milliamps. On the other hand, the RF power for superconducting cavities is dominated by the beam even for very small currents of less than 0.1 mA. The ratio between the total RF power of normal conducting and superconducting cavities including the beam is decreasing for higher beam currents. For very high currents such as 100 mA, the ratio decreases to a factor below two.

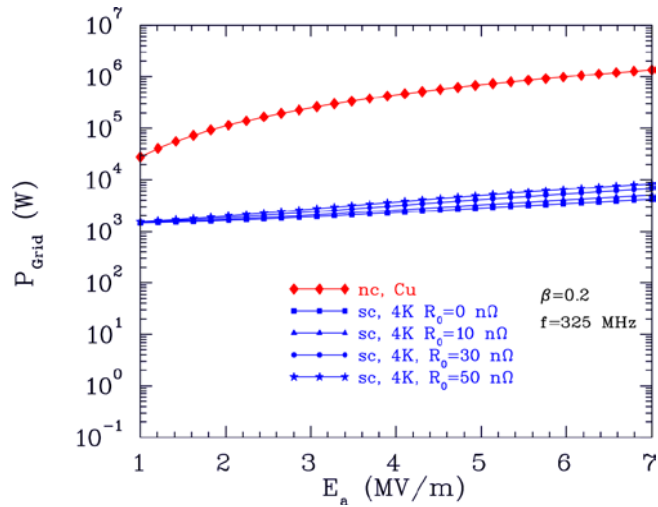

**Fig. 8:** Grid power without beam as a function of the accelerating gradient $E_a$, duty factor 100 %. Normal conducting cavity: $f$ = 325 MHz, $Z_{eff}$ = 60 MΩ/m, $\beta$ = 0.2, $L$ = 1 m. Superconducting cavity: $f$ = 325 MHz, $\beta$ = 0.2, $R/Q$ = 3000 Ω, $T$ = 4.2 K, $G$ = 55 Ω, static losses = 5 W.

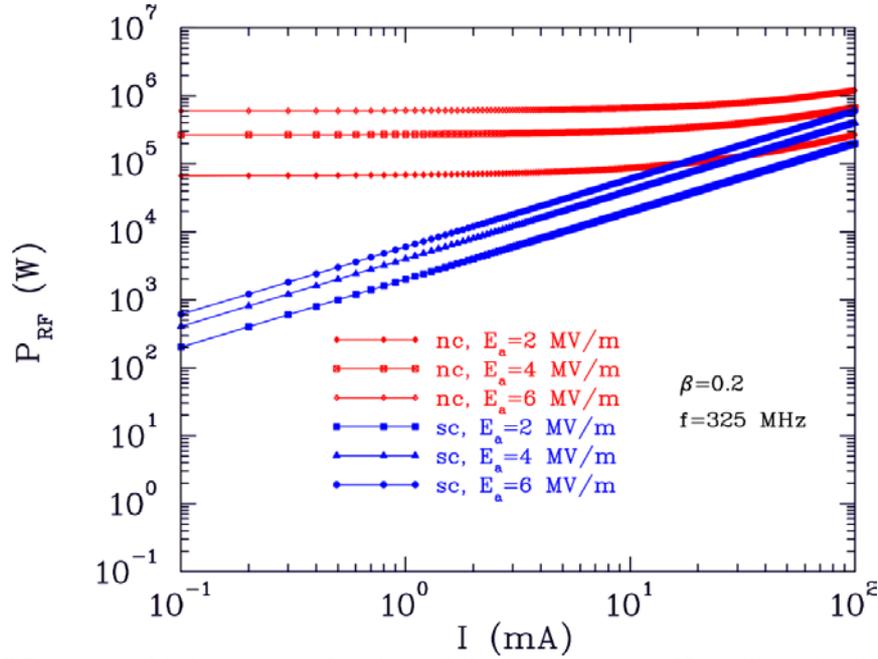

**Fig. 9:** RF power with beam as a function of the beam current. Normal conducting cavity: $f = 325$ MHz, $Z_{eff} = 60$ MΩ/m, $\beta = 0.2$, $L = 1$ m. Superconducting cavity: $f = 325$ MHz, $\beta = 0.2$, $R/Q = 3000$ Ω, $T = 4.2$ K, $G = 55$ Ω.

If superconducting cavities are operated at low duty cycles, the static losses can play an important role regarding the required grid power. Figure 10 shows the grid power as function of the beam current for different duty cycles. The higher the duty cycle the faster the grid power is dominated by the beam power in the case of superconducting cavities. The lower the duty cycle and the higher the current, the more similar is the grid power between normal conducting and superconducting cavities.

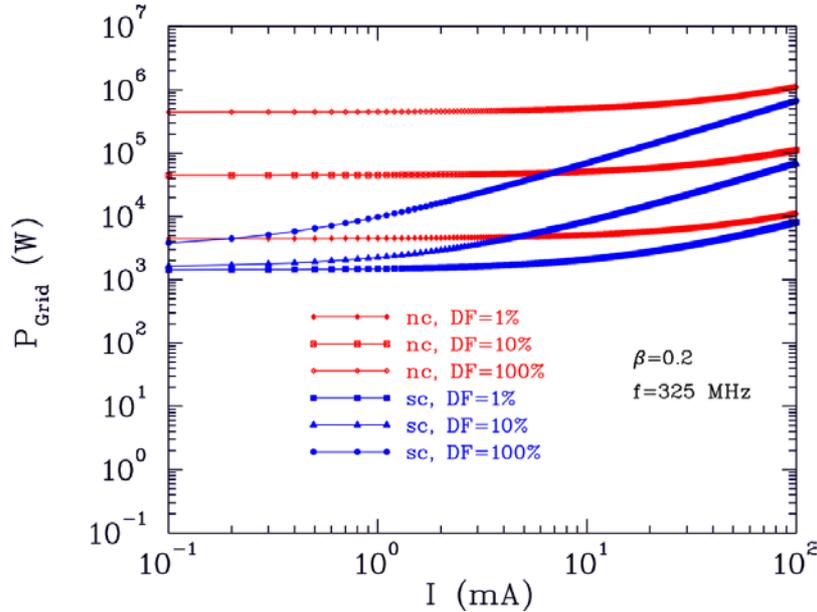

**Fig. 10:** Grid power with beam as a function of the beam current. Normal conducting cavity: $f = 325$ MHz, $Z_{eff} = 60$ MΩ/m, $\beta = 0.2$, $L = 1$ m. Superconducting cavity: $f = 325$ MHz, $\beta = 0.2$, $R/Q = 3000$ Ω, $T = 4.2$ K, $G = 55$ Ω, static losses = 5 W, $R_s = 30$ nΩ.

For low duty cycle machines with low beam power, the grid power can be lower for normal conducting compared with superconducting operation. Figure 11 shows the grid power without beam as a function of the duty cycle. For the superconducting cavity different values for static losses and residual resistance have been assumed. In this example ($\beta = 0.2$, $f = 325$ MHz) the break-even takes place at a duty factor of about 1 %. In particular, for lower beam energies there is often a favour for normal conducting cavities because of the existence of cavities with very high shunt impedance such as IH cavities [11].

Figure 12 shows the grid power as function of the duty factor for different beam currents ($E_a = 4$ MV/m). The power for normal conducting cavities increases linearly with the duty factor while for superconducting cavities this is only true for high beam currents and high duty factor. At lower duty factor and beam current, the power is dominated by the static losses.

Another interesting question is about the efficiency of the cavity to convert grid power into beam power (Fig. 13). As expected, the ratio of average beam power and grid power as a function of the duty cycle is constant for normal conducting cavities. It increases for higher beam current. Owing to the static losses of superconducting cavities the ratio increases for these cavities. For higher beam currents ($I > 10$ mA) the curves reach a plateau at a value which represents the efficiency of the amplifier.

Figure 14 shows the ratio of average beam power and grid power as function of the beam current. For superconducting cavities the curves reach the amplifier efficiency earlier for higher duty factor because of the higher average beam power. Normal conducting cavities typically reach only 50 % of the amplifier efficiency at reasonable beam currents.

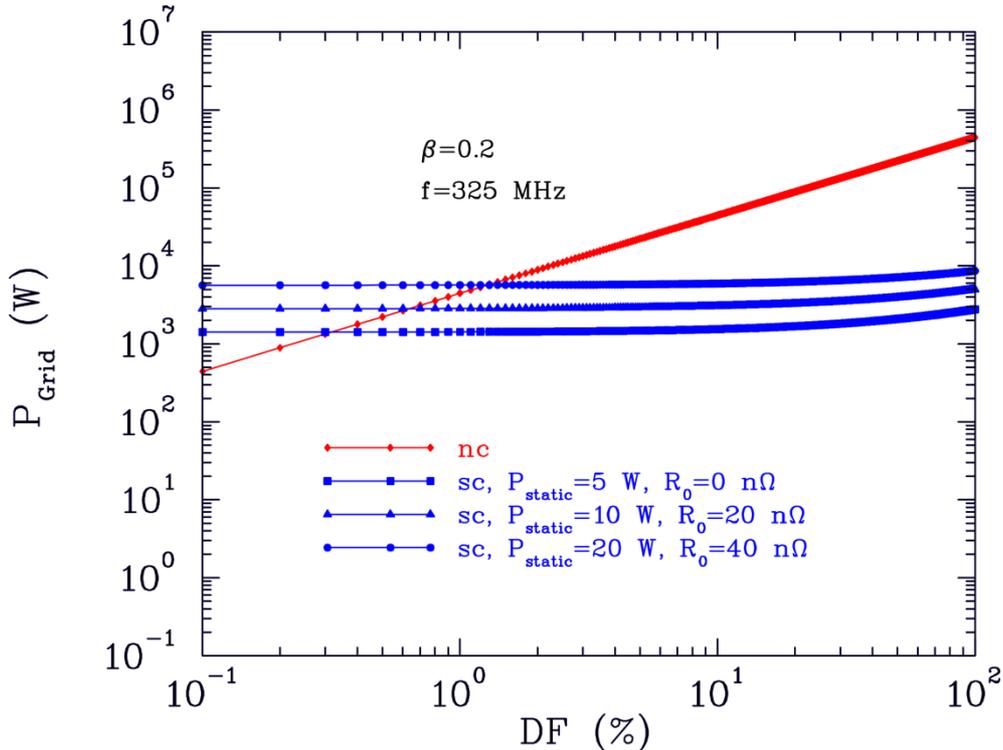

**Fig. 11:** Grid power without beam as a function of the duty factor, $E_a = 4$ MV/m. Normal conducting cavity: $f = 325$ MHz, $Z_{eff} = 60$ MΩ/m, $\beta = 0.2$, $L = 1$ m. Superconducting cavity: $f = 325$ MHz, $\beta = 0.2$, $R/Q = 3000$ Ω, $T = 4.2$ K, $G = 55$ Ω.

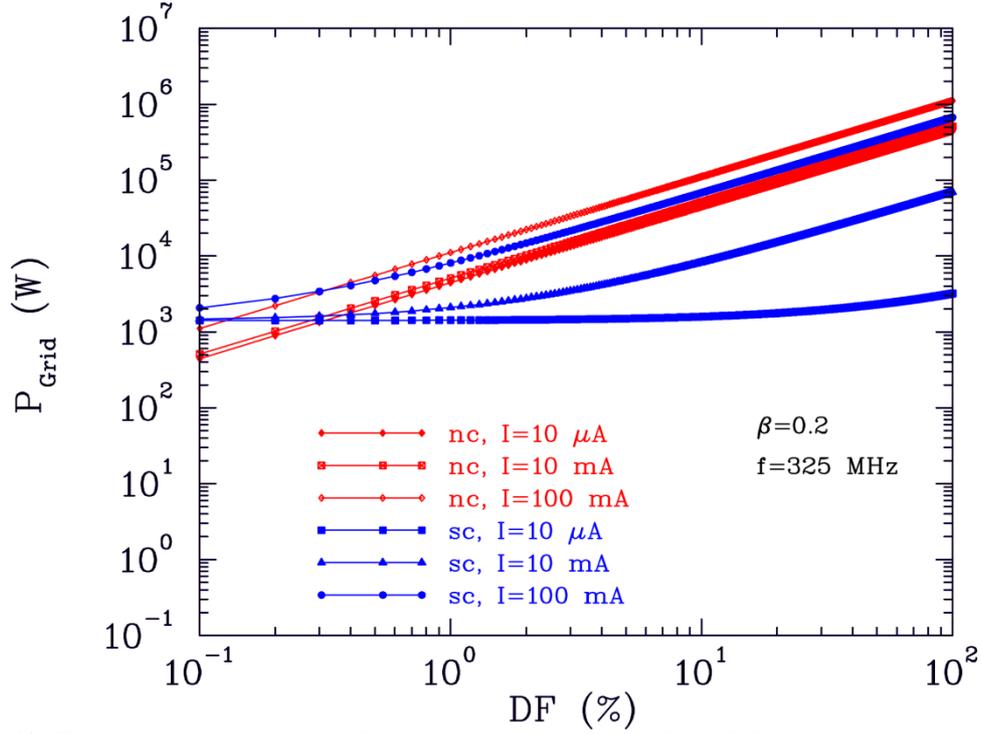

**Fig. 12:** Grid power with beam as a function of the duty factor, $E_a$ = 4 MV/m. Normal conducting cavity: $f$ = 325 MHz, $Z_{eff}$ = 60 MΩ/m, β = 0.2, $L$ = 1 m. Superconducting cavity: $f$ = 325 MHz, β = 0.2, $R/Q$ = 3000 Ω, $T$ = 4.2 K, $G$ = 55 Ω, static losses = 5 W, $R_s$ = 30 nΩ.

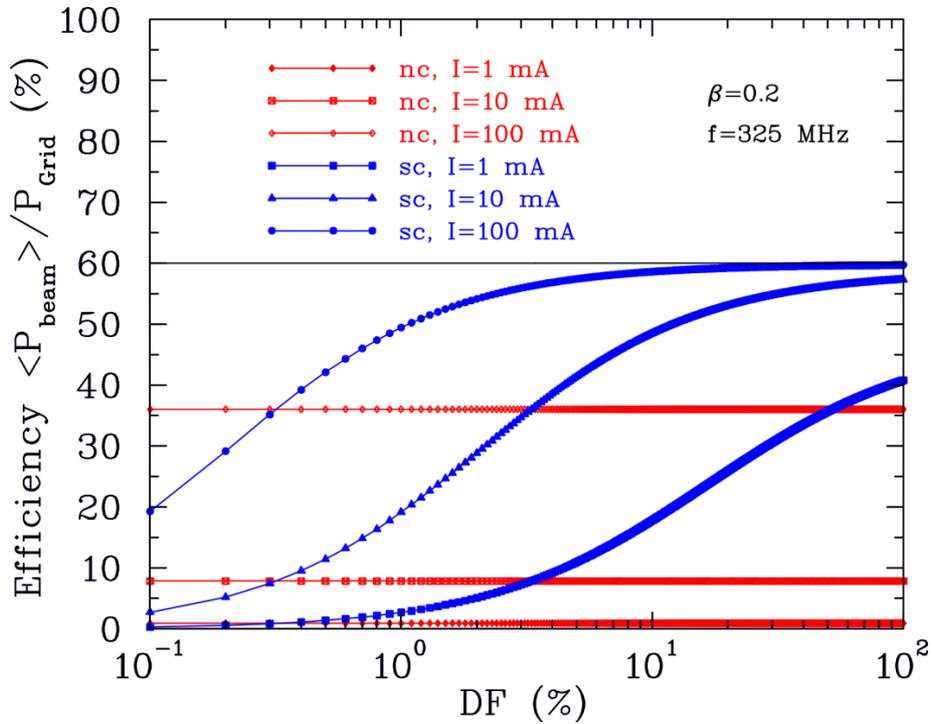

**Fig. 13:** Efficiency $P_{beam}/P_{Grid}$ with beam as a function of the duty factor, $E_a$ = 4 MV/m. Normal conducting cavity: $f$ = 325 MHz, $Z_{eff}$ = 60 MΩ/m, β = 0.2, $L$ = 1 m. Superconducting cavity: $f$ = 325 MHz, β = 0.2, $R/Q$ = 3000 Ω, $T$ = 4.2 K, $G$ = 55 Ω, static losses = 5 W, $R_s$ = 30 nΩ.

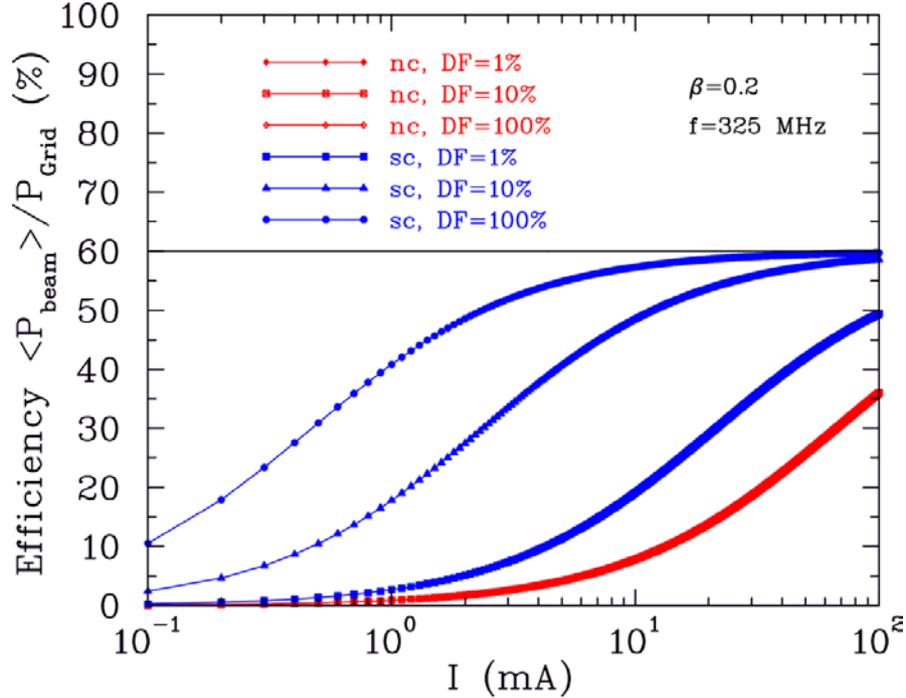

**Fig. 14:** Efficiency $P_{beam}/P_{Grid}$ with beam as a function of the beam current, $E_a = 4$ MV/m. Normal conducting cavity: $f = 325$ MHz, $Z_{eff} = 60$ MΩ/m, $\beta = 0.2$, $L = 1$ m. Superconducting cavity: $f = 325$ MHz, $\beta = 0.2$, $R/Q = 3000$ Ω, $T = 4.2$ K, $G = 55$ Ω, static losses = 5 W, $R_s = 30$ nΩ.

## 5 Limitations of normal conducting and superconducting cavities

During testing and operating normal conducting and superconducting cavities we can observe different phenomena which are sometimes real limitations. In principle, these phenomena exist in all cavities independent of whether normal conducting or superconducting, but the severity of the problems can be very much different for normal conducting and superconducting cavities.

The most serious limitation of normal conducting cavities is the thermal load for high duty cycle operation which becomes equal to the RF losses for cw operation. Above a certain thermal load it is no longer possible to remove the heat from the cavity walls without losing performance or reliability. A typical upper limit is about 100 kW/m for normal conducting cavities, but it can be lower or higher depending on the used material (thermal conductivity), the geometry and the power density distribution.

In the case of superconducting cavities the thermal load is less important because of the low-power losses. Locally heated superconducting material has a higher surface resistance, however, resulting in even more power dissipation. This could finally lead to thermal breakdown of superconductivity (quench) above a specific field level. Sometimes there are small normal conducting inclusions within the superconducting material. In these typically sub-millimetre size defects we have significantly higher power densities. Below a power threshold, the superconducting material can conduct the additional heat while remaining superconducting. Above this threshold, the surrounding superconductor reaches the critical temperature leading to a quench. In the case of superconducting cavities there is a fundamental limitation for the maximum magnetic surface field in the cavity. Above a material specific magnetic field (200 mT for Nb) there is a breakdown of superconductivity. In reality only very few elliptical cavities have reached field levels close to this fundamental limit. Nevertheless, an important design issue for superconducting cavities is the minimization of the magnetic peak fields.

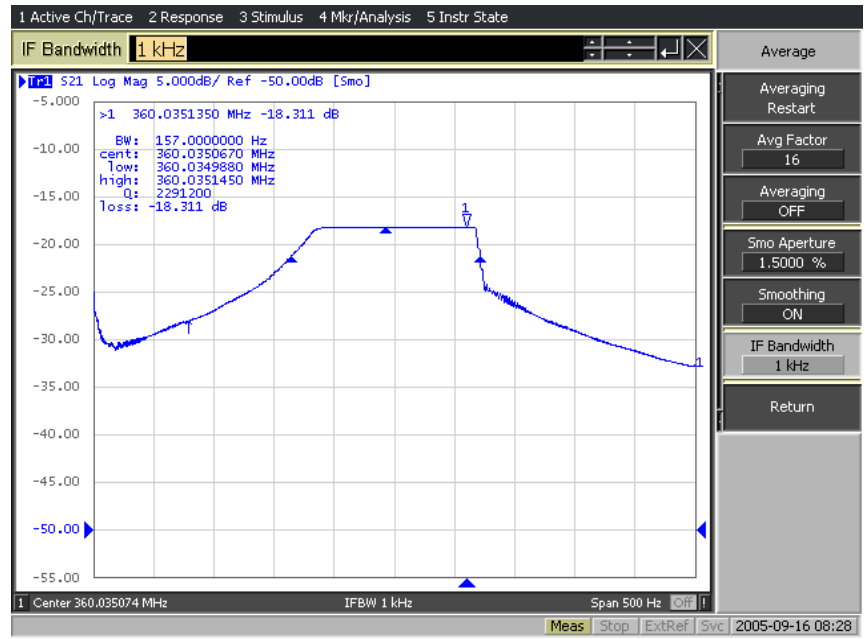

**Fig. 15:** Multipacting during the test of a superconducting cavity. The frequency has been swept over the resonance. Above a certain threshold (multipacting barrier) the field level remains constant.

A common problem of RF cavities can be multipacting. It is a rapid growing electron avalanche typically in the low-electric-field region. Low-energy electrons with energies of a few hundred electronvolts can hit the cavity wall and thus create secondary electrons. These electrons are accelerated and bent in the time-varying electromagnetic fields. They can hit the walls again leading to an increasing number of electrons if certain circumstances (electron energy, field distribution, frequency, field level) are present. If multipacting occurs and a so-called multipacting barrier has been reached it is not possible to increase the field level in the cavity because the stored energy and additional power is used to create the electron avalanche.

By sweeping the frequency over the resonance a flat top appears. Figure 15 shows the resonance curve with a flat top during the test of a superconducting cavity [12]. In most cases it is possible to overcome these barriers (soft barriers) with different conditioning methods. The shorter the time which is available for the avalanche creation the easier is the conditioning. This makes it clear why superconducting cavities typically suffer more from multipacting. The rise time of the fields in superconducting cavities is normally much longer than in normal conducting cavities. Stronger coupling of the cavities can decrease the rise time and shorten the conditioning time. Further information regarding multipacting can be found in Ref. [13].

The $Q$ value is determined by the ratio of stored energy and cavity loss. Because both quantities increase quadratically with the field level, the $Q$ value should be independent of the gradient, but most of the superconducting cavities show a significant decrease of the $Q$ value at higher gradients. Figure 16 shows a typical measurement of the $Q$ value as a function of the gradient [14].

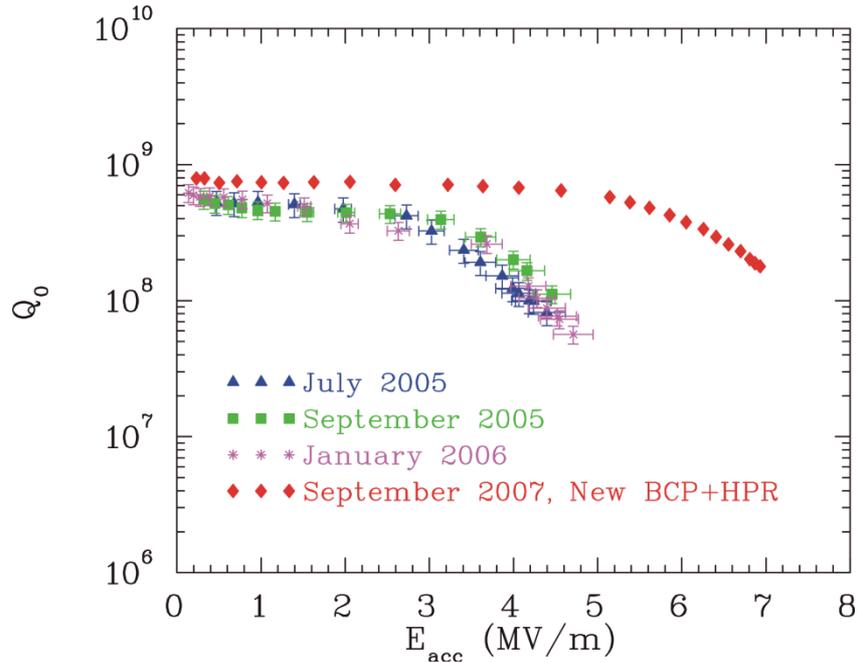

**Fig. 16:** Measured *Q* value of a superconducting cavity as function of the gradient. At higher field level the *Q* value decreases significantly [12].

A decreasing *Q* value means that the cavity power is increasing faster than the stored energy because of non-Ohmic losses. The most common reason for these additional losses is field emission. Field enhancements in the high-electric-field region can lead to emission of electrons which are then accelerated in the electric fields. Additional power is required resulting in the lower *Q* value. Electrons hitting the cavity are decelerated leading to the emission of X-rays. This can occur in normal conducting as well as in superconducting cavities. If the dissipated power is too high a thermal breakdown of superconductivity can happen. Typically superconducting cavities suffer by far more from field emission than normal conducting cavities because of the low dissipated power. Therefore, it is necessary to reduce the risk of field emission by special surface preparation methods (buffered chemical polishing [BCP], high-pressure rinsing [HPR]) and following assembly in a clean room. To reduce the risk of field emission one important design issue for superconducting cavities is to reduce electric peak fields and the ratio between gradient and peak field $E_a/E_p$.

Another possible problem is microphonics which affects mostly superconducting cavities. A cavity can oscillate mechanically in the acoustic frequency regime due to external excitations. These oscillations leading to a deformation of the cavity and finally to shifts of the RF frequency which can be much larger than the bandwidth of the cavity. There are different options to overcome this problem. We can use mechanical stiffeners to reduce the amplitude of the vibrations and to increase their frequency which makes excitation less probable. Another possibility is the use of fast-acting tuners mostly based on piezo-crystals. A stronger RF coupling (over-coupling) leads to a broader resonance curve for the price of additional RF and reflected power.

The last problem which should be mentioned is Lorentz force detuning (LFD). The electromagnetic fields inside a cavity create a pressure resulting in a field-level-dependent deformation of the cavity. This deformation leads to a detuning of the cavity ($\Delta f < 0$) which is proportional to the stored energy and the square of the field level, respectively. For cw operated superconducting cavities or normal conducting cavities in general LFD is normally not a problem, but for pulsed thin-walled superconducting cavities the detuning can be significant larger than the cavity bandwidth. In this case often fast tuners with feed-forward systems are used.

# 6     Case studies

In this section two linac projects are described and compared, each with normal conducting and superconducting technology. The first project is the 70 MeV proton injector for the Facility for Anti-Proton and Ion Research (FAIR) which is a typical normal conducting linac because of the high beam current and low duty cycle. The second example is the superconducting cw linac at GSI which with a duty factor of 100 % and low beam current.

## 6.1    NC: FAIR proton linac

The 70 MeV proton injector for the FAIR is needed to fulfil the requirements for the experimental program with respect to beam current and beam pulse structure [15]. It is the injector for the heavy ion synchrotron SIS100. It accelerates a peak proton current of up to 70 mA to a final energy of 70 MeV. The duty factor is below 0.1 %. For this project, six normal conducting CH-drift tube cavities are used [16]. The linac length is about 25 m. Owing to the very low duty cycle and the available 325 MHz, 3 MW klystrons relatively high gradients between 3 MV/m and 7 MV/m are used. Taking only the power to operate the cavities into account about 15 kW of grid power is needed. In the case of a superconducting version 14 CH cavities with a gradient of 5 MV/m would be required. This number is higher than for the normal conducting linac because it is not possible to integrate the magnetic focusing elements inside the superconducting cavities. The overall length is about the same. Assuming static losses of 10 W/m and 4 K operation we need about 100 kW of plug power to operate the cavities. Although this is much higher than for a normal conducting linac the main disadvantage of the superconducting solution would be the significant higher capital costs for the cavities, cryomodules, klystrons and cryogenic plant. Table 3 summarizes the main parameters of a normal conducting and a superconducting FAIR proton linac. Figure 17 shows the schematic layout of the normal conducting linac using CH-cavities and Fig. 18 shows the superconducting linac.

**Table 3**: Comparison of a normal conducting and a superconducting FAIR proton injector

| Parameter | Normal conducting | Superconducting |
| --- | --- | --- |
| Particles | Protons | Protons |
| Frequency (MHz) | 325 | 325 |
| Gradient (MV/m) | 3–7 | 5 |
| Energy (MeV) | 70 | 70 |
| Beam current (mA) | 70 | 70 |
| RF structure | nc CH | sc CH |
| Linac length (m) | 25 | 25 |
| Pulse length RF (µs) | 100 | 200 |
| Pulse length beam (µs) | 36 | 36 |
| Repetition rate (Hz) | 4 | 4 |
| Duty factor (%) | <0.1 | <0.1 |
| Number of cavities | 6 | 14 |
| P klystron (kW) | 3000 | 500 |
| Grid power (kW) | 15 | 100 |
| $P_{beam}/P_{Grid}$ (%) | 4.5 | 0.7 |

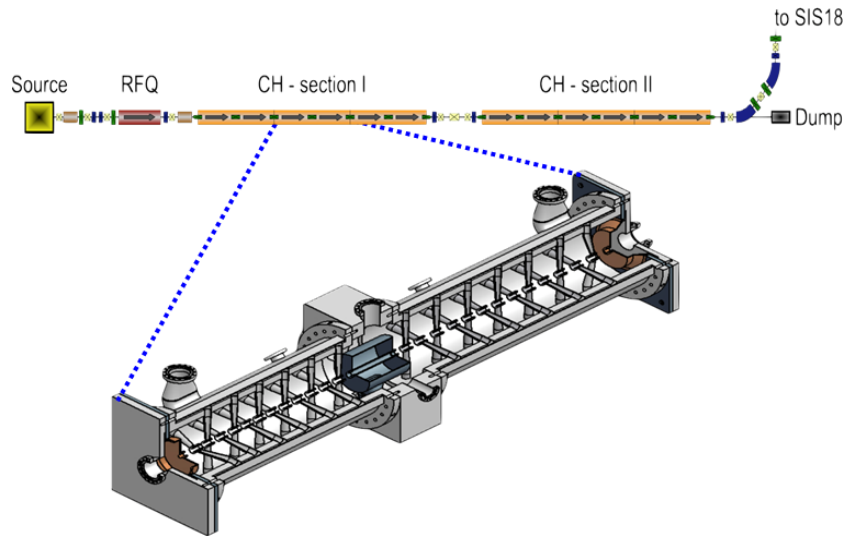

**Fig. 17:** Schematic layout of the normal conducting 70 MeV FAIR proton linac using CH cavities

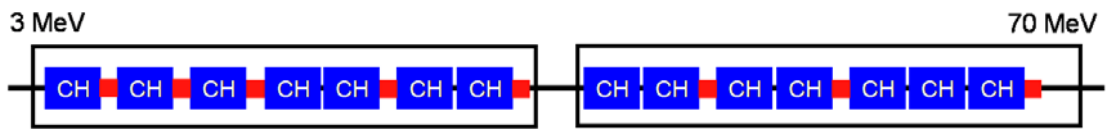

**Fig. 18:** In the case of a superconducting FAIR proton linac, more cavities and klystron would be required. The grid power would be significantly higher because of the low duty cycle and high beam current.

### 6.2  SC: GSI cw SHE Linac

The superconducting cw Super Heavy Elements (SHE) linac is a heavy ion linac operated cw [10]. It accelerates heavy ions with an *A*/*q* ratio of 6 from 1.4 AMeV to energies of up to 7.3 AMeV. Owing to the cw operation and the relatively low beam current of 100 µA, this linac is predestined to be operated in the superconducting mode. The front end is the existing High Charge State Injector at GSI [17].

The superconducting version consists of 9 CH cavities with superconducting solenoids for transverse focusing. For each cavity a 5 kW amplifier is foreseen. The total heat load at 4 K is estimated to be 500 W leading to about 200 kW of grid power. The total grid power with RF amplifiers is about 275 kW. Figure 19 shows the schematic layout of the superconducting version. The linac after the injector is about 12 m long (Fig. 19).

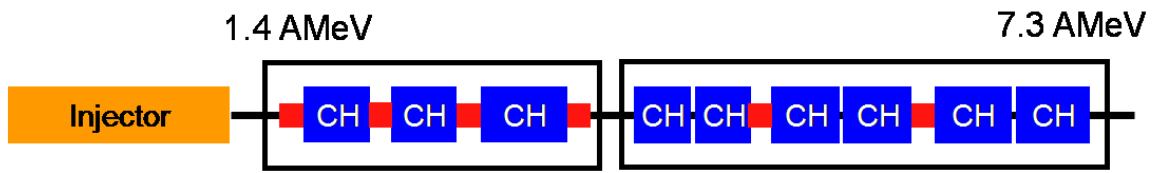

**Fig. 19:** Schematic layout of the superconducting cw linac at GSI. Nine CH cavities with gradients of 5 MV/m are foreseen.

In the case of a normal conducting linac we have to reduce the gradient from 5 MV/m to values between 1.5 MV/m and 2 MV/m. In total at least 14 normal conducting cavities (here IH-cavities) would be needed (Fig. 20). The required RF power for these cavities would be 900 kW resulting in a grid power of 1500 kW. This leads to 7 500 000 kW h additional energy per year. The linac length would increase by more than a factor of 2.5. Table 4 summarizes the main parameters of a normal conducting and a superconducting cw SHE linac.

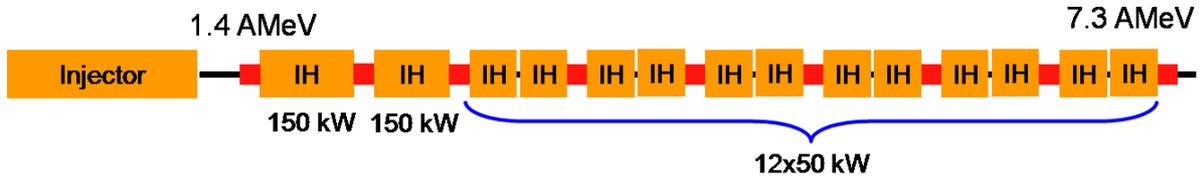

**Fig. 20:** In the case of a normal conducting cw linac at GSI using efficient IH cavities the required plug power would be five times the amount of a superconducting solution

Table 4: Comparison of a normal conducting and a superconducting cw SHE-linac at GSI

| Parameter | Normal conducting | Superconducting |
| --- | --- | --- |
| Particles | Heavy ions | Heavy ions |
| Frequency (MHz) | 217 | 217 |
| Gradient (MV/m) | 1.5–2.0 | 5.1 |
| Energy (AMeV) | 7.3 | 7.3 |
| $A/q$ | 6 | 6 |
| Beam current (μA) | 100 | 100 |
| RF structure | nc IH | sc CH |
| Linac length (m) | 30 | 12 |
| Duty factor (%) | 100 | 100 |
| Number of cavities | 14 | 9 |
| Amplifier (kW) | 50–150 | 5 |
| Grid power (kW) | 1500 | 275 |
| $P_{beam}/P_{Grid}$ (%) | 0.24 | 1.13 |

## 7   Summary

Each linac project has to make the choice which is the best-suited technology for the specific application. The main issues are capital costs, operating costs (power), technical risk and reliability. During the last few decades the transition energy between normal conducting and superconducting technology decreased significantly. The main reasons for this are lower operation costs and especially the availability of suitable superconducting RF structures in the low- and medium-energy range such as quarter-wave resonators, half-wave resonators and CH cavities.

Normally the shunt impedance is higher at low energies. There are very efficient low-energy drift tube structures such as IH cavities available. Above 100 AMeV or 200 AMeV normally superconducting cavities are the best choice even for machines with lower duty cycle because there are no real efficient normal conducting structures available. On the other side superconducting elliptical cavities can reach high gradients ($E_a > 10$ MV/m) resulting in a much shorter linac.

In general, normal conducting cavities are more favourable at lower energies with high beam current and low duty cycle. For superconducting cavities the opposite is valid (Fig. 21). Both technologies have their advantages and disadvantages (Fig. 22). Superconducting technology requires a cryogenic plant with associated helium distribution. The superconducting cavities are very sensitive against contaminations, pressure variations and vibrations and need in certain circumstances fast tuner systems. On the other hand, they can be operated very reliably even with cw operation.

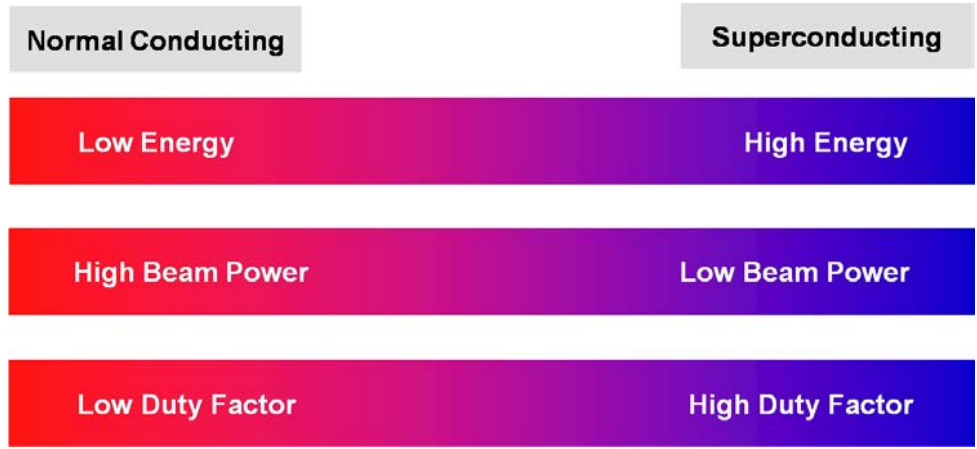

**Fig. 21:** Regimes of normal conducting and superconducting cavities

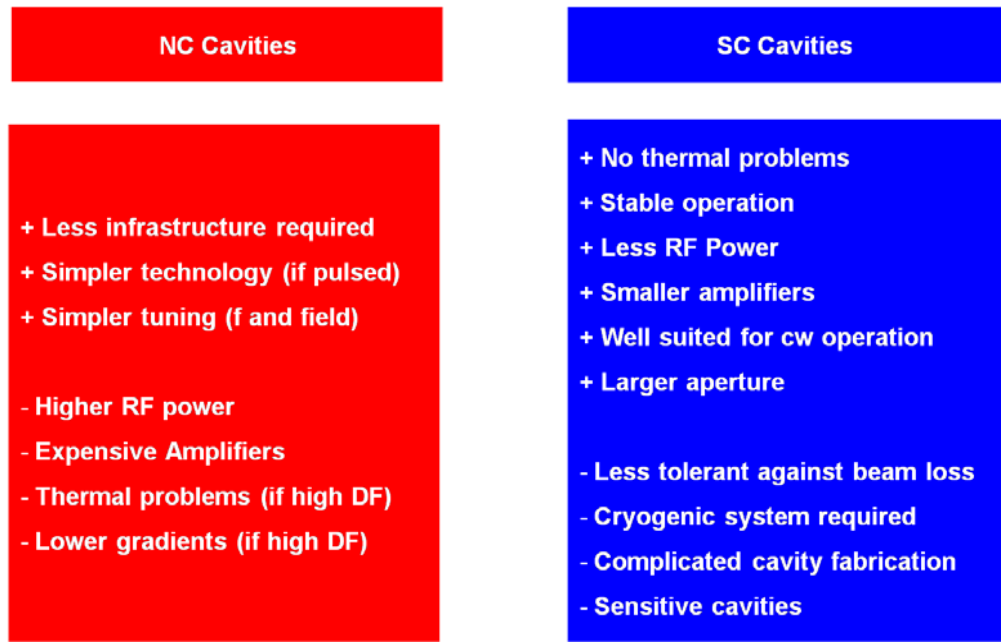

**Fig. 22:** Advantages and disadvantages of normal conducting and superconducting cavities